\documentclass[aps,prd,preprintnumbers,showpacs,showkeys,nofootinbib,
superscriptaddress,fleqn,floatfix,tightenlines, 10pt]{revtex4-1}
\usepackage{amsmath,amsfonts,amssymb,amscd,amsxtra,amsthm}
\usepackage{graphicx}  
\usepackage{epstopdf}
\usepackage{dcolumn}  
\usepackage{bm}          
\usepackage{slashed}
\usepackage[utf8]{inputenc} 
\usepackage[normalem]{ulem} 
\usepackage[dvipsnames]{xcolor} 
\usepackage{array}
\usepackage{slashed}
\renewcommand\sout{\bgroup \color{red} \ULdepth=-.5ex \ULset}
\newcommand{\com}[1]{{\sf\color[rgb]{0,0,1}{#1}}}

\usepackage[caption = false]{subfig}
\usepackage{multirow}

\begin{document} 
\preprint{INHA-NTG-10/2018}
\title{Vector and Axial-vector form factors in radiative kaon decay
  and flavor SU(3) symmetry breaking}
\author{Sang-In Shim}
\email[E-mail: ]{shimsang@rcnp.osaka-u.ac.jp}
\affiliation{Research Center for Nuclear Physics (RCNP),
Osaka University, Ibaraki, Osaka, 567-0047, Japan}
\author{Atsushi Hosaka}
\email[E-mail: ]{hosaka@rcnp.osaka-u.ac.jp}
\affiliation{Research Center for Nuclear Physics (RCNP),
Osaka University, Ibaraki, Osaka, 567-0047, Japan}
\affiliation{Advanced Science Research Center, Japan Atomic Energy
  Agency, Shirakata, Tokai, Ibaraki, 319-1195, Japan}
\author{Hyun-Chul Kim}
\email[E-mail: ]{hchkim@inha.ac.kr}
\affiliation{Department of Physics, Inha University, Incheon 22212,
Republic of Korea}
\affiliation{Advanced Science Research Center, Japan Atomic Energy
  Agency, Shirakata, Tokai, Ibaraki, 319-1195, Japan}
\affiliation{School of Physics, Korea Institute for Advanced Study 
  (KIAS),\\ Seoul 02455, Republic of Korea}
\date{\today}
\begin{abstract}
We study the vector and axial-vector form factors of radiative
kaon decay within the framework of the gauged nonlocal effective
chiral action from the instanton vacuum, focusing on the effects of
flavor SU(3) symmetry breaking. The general tendency of the results
are rather similar to those of radiative pion decays: The nonlocal
contributions make the results of the vector form factor increased by
about $20\,\%$, whereas they reduce those of the axial-vector form
factor by almost \com{$50\,\%$}. Suppressing the prefactors consisting of 
the kaon mass and the pion decay constant, we scrutinize how the kaon 
form factors undergo changes as the mass of the strange current quark 
is varied. Those related to the vector and second axial-vector form 
factors tend to decrease monotonically as the strange quark mass 
increases, whereas that for the axial-vector form factor decreases. 
When $K\to e\nu\gamma$ decay is considered, both the results of the 
vector and axial-vector form factors at the zero momentum transfer are 
in good agreement with the experimental data. The results are also 
compared with those from chiral perturbation theory to $p^6$ order.
\end{abstract}
\maketitle
\section{Introduction}
Radiative kaon decay ($K_{l2\gamma}$) provides essential information
on the structure of the kaon. Though the structure of the radiative
kaon form factors is very similar to that of the pion decay, the
effects of flavor $\mathrm{SU}(3)$ symmetry breaking, which arises
from the current mass of the strange quark, makes the kaon
distinguished from the pion.  As in the case of the pion, the
radiative decay amplitude for the $K^+\to l^+ + \nu_l +\gamma$ can be
decomposed into two parts, i.e., the structure-dependent (SD) part,
and the inner Bremsstrahlung (IB) one or the QED
corrections~\cite{Vaks:1958,Bludman:1960, Bryman:1982et,
  Neville:1961zz, Kanazawa:1961, Cirigliano:2011ny, PDG}. The
decay rate is given by the squared modulus of the amplitude, so that
it consists of three different terms, that is, the IB one, the mixed one,
and the pure SD one. The IB and mixed terms are proportional to the
squared ratio of the lepton and kaon masses, i.e., $(m_l/m_K)^2$,
which is called the helicity suppression factor, the radiative kaon
decay to the electron ($K_{e2\gamma}$) is governed by the SD term. On 
the other hand, $K_{\mu2\gamma}$ is sensitive to the IB and mixed
terms. Neville~\cite{Neville:1961zz} proposed that by choosing the
angle between the neutrino and the photon with the helicities of both
the lepton and neutrino fixed one could measure the vector and
axial-vector form factors of $K_{l2\gamma}$. 

The suggestion of Ref.~\cite{Neville:1961zz} being considered, the radiative kaon decay
$K_{e2\gamma}$ was measured several decades ago~\cite{Heard:1974kk,
  Heintze:1976qf, Heintze:1977kk}. Heintze et
al.~\cite{Heintze:1977kk} extracted the following results:
$|F_V+F_A|=0.148\pm0.010$ and $|F_V-F_A|<0.49$ in the standard
notation~\cite{PDG}, where $F_V$ and $F_A$ denote the vector and
axial-vector form factors of the radiative kaon decay.
In Refs.~\cite{Akiba:1985zh, Barmin:1987gp, Demidov:1989gn}, the 
$K_{\mu 2\gamma}$ decay was experimentally studied but the form
factors were not extracted, since they investigated mainly the
IB-dominant region. The E787 Collaboration~\cite{Adler:2000vk}
performed the first measurement of an SD component in the radiative
kaon decay $K_{\mu2\gamma}$ and extracted
$|F_V+F_A|=0.165\pm 0.007 \,(\mathrm{stat})\pm 0.011\,(\mathrm{syst})$
and the limit $-0.04<F_V-F_A<0.24$ at $90\,\%$ confidence level.
The E865 experiment at the Brookhaven National Laboratory
(BNL)~\cite{Poblaguev:2002ug} studied experimentally the kaon
radiative decays $K^+\to \mu^+ \nu e^+e^-$ and $K^+\to e^+ \nu
e^+e^-$, where the photon is in a virtual state. When $\gamma$ is
virtual, yet an additional form factor $R$ is involved. Though the
analysis of Ref. ~\cite{Poblaguev:2002ug} inevitably 
contains the model dependence, the results were obtained to be 
$F_V=0.112\pm 0.015\pm 0.010\pm 0.003\,(\mathrm{model})$, 
$F_A=0.035\pm 0.014\pm 0.013\pm 0.003$, and $R=0.227\pm0.013\pm
0.010\pm 0.009$, when both the data of radiative decays $K^+\to \mu^+
\nu e^+e^-$ and $K^+\to e^+ \nu e^+e^-$ were combined. The KLOE
Collaboration~\cite{Ambrosino:2009aa} 
measured the ratio $\Gamma(K_{e2(\gamma)}/\Gamma(K_{\mu2(\gamma)})$
and obtained also the sum of the vector and axial-vector form factors
as $F_V+F_A=0.125\pm 0.007\pm 0.001$. Some years ago, ISTRA+
Collaboration~\cite{Duk:2010bs} reported the extraction of the form
factors from the $K^-\to \mu^-\nu \gamma$ decay: $F_V-F_A=0.21 \pm
0.04\pm 0.04$ with the sign also determined. We want to mention that
the value of $F_V-F_A$ extracted from the E865 experiment is in
disagreement with that from the ISTRA+ Collaboration.

The vector and axial-vector form factors for the radiative kaon decay
were studied in chiral perturbation theory
($\chi$PT)~\cite{Gasser:1984gg, Donoghue:1989si, Bijnens:1992mk,
  Bijnens:1992en, Geng:2003mt}, since the experimental data on the
axial-vector form factors can be used to determine a part of the
low-energy constants (LECs) that reflect certain features of
nonperturbative quark-gluon dynamics. Geng et al. examined 
the form factors for the radiative kaon decay to order
$\mathcal{O}(p^6)$~\cite{Geng:2003mt}. The analysis of $K_{e2\gamma}$
was also carried out within the light-front quark
model~\cite{Chen:2007bv} in which the dependence of $F_V$ and $F_A$ on
the momentum transfer squared was presented. 
ISTRA+ Collaboration ~\cite{Duk:2010bs} showed explicitly that the
results of $\mathcal{O}(p^6)$ $\chi$PT is found to  
be slightly away from the $3\,\sigma$ ellipse. It is also interesting 
to see that the radiative kaon decay could be used for understanding a
possible new physics such as the massless dark
photon~\cite{Carlson:2013mya, Fabbrichesi:2017vma}.  
The radiative kaon decay provides yet another theoretically important
aspect on the effects of flavor $\mathrm{SU}(3)$ symmetry breaking.
The structure of the radiative transition amplitude of the kaon is the 
same as that of the pion except for the difference of their masses. It
indicates that by suppressing the kinematical factors in the
expressions of the form factors one can scrutinize how the current
quark mass comes into play in describing the radiative kaon and pion
decays.  

In this work, we investigate the vector and axial-vector form factors
for the radiative kaon decay within the framework of 
the gauged nonlocal effective chiral action (the gauged E$\chi$A)
 from the instanton
vacuum~\cite{Diakonov:1985eg, Diakonov:2002fq, Musakhanov:2002xa,
  Kim:2004hd, Kim:2005jc, Goeke:2007nc, Goeke:2007bj}, emphasizing the
effects of the flavor $\mathrm{SU}(3)$ symmetry breaking. The
instanton vacuum offers a natural realization of the spontaneous
breakdown of chiral symmetry (SB$\chi$S). Consequently, the kaon
appears as a pseudo-Nambu-Goldstone boson like the pion and the mass
of the quark is dynamically generated. This dynamical quark mass 
is momentum-dependent, which is originated from the fermionic zero
modes in the instanton background. There are two  
parameters characterizing the diluteness of the instanton liquid,
namely, the average instanton size $\bar{\rho} \approx 1/3$ fm and
average interinstanton distance $\bar{R} \approx 1$ fm. In particular,
the average size of instantons is considered as a normalization point
equal to $\bar{\rho}^{-1} \approx 0.6$ GeV. It implies that the
results of the scale-dependent quantities from the model 
can be easily compared with those from other theoretical 
framework such as $\chi$PT and lattice QCD. These values of the
$\bar{\rho}$ and $\bar{R}$ were determined many years ago 
by a variational method~\cite{Diakonov:1985eg,Diakonov:2002fq} as  
well as phenomenologically~\cite{Shuryak:1981ff,Schafer:1996wv}. Note
that these values were also confirmed within lattice
QCD~\cite{Chu:1994vi, Negele:1998ev, DeGrand:2001tm}.  
Reference~\cite{Cristoforetti:2006ar} simulated the QCD vacuum in the 
interacting instanton liquid model and derived $\bar{\rho}\approx
0.32\,\mathrm{fm}$ and $\bar{R}\approx 0.76\,\mathrm{fm}$ with the
finite current quark mass taken into account.  

When the finite current quark mass is considered, one needs to modify
the E$\chi$A derived by Diakonov and Petrov~\cite{Diakonov:1985eg}.
The modification was performed by
Musakhanov~\cite{Musakhanov:1998wp, Musakhanov:2001pc,
  Musakhanov:2002vu}. In particular, 
it is essential to include the current quark mass in the E$\chi$A,
when one deals with properties of the kaon. It was shown in
Ref.~\cite{Nam:2006ng} that the improved E$\chi$A explained very well
the dependence of the quark and gluon condensates on the current quark 
mass. Furthermore, the momentum-dependent dynamical quark mass 
brings about the breakdown of the Ward-Takahashi (WT) 
identities, that is, the current nonconservation~\cite{Chretien:1954we,
Pobylitsa:1989uq, Bowler:1994ir, Musakhanov:1996cv}. 
Musakhanov and Kim~\cite{Musakhanov:2002xa, Kim:2004hd} showed how 
the gauge-invariant E$\chi$A can be constructed, which satisfies the
WT identities. By using 
the gauged E$\chi$A, various properties of the $\pi$ and $K$ mesons
such as electromagnetic form factors~\cite{Nam:2007gf}, kaon
semileptonic decay form factors~\cite{Nam:2007fx}, distribution
amplitudes~\cite{Nam:2006sx} and the pion weak form
factors~\cite{Shim:2017wcq} have been studied. We will use this gauged
E$\chi$A in the present work to investigate the weak vector and
axial-vector form factors for the radiative kaon decay which is
similar to the case of pion by some of the authors~\cite{Shim:2017wcq}. 

The present work is outlined as follows: In Section
II, we define one vector and two axial-vector form factors for the
radiative kaon decay. In Section III, we explain the gauged
E$\chi$A, focusing on the radiative kaon decay. In Section IV, we
compute the vector and axial-vector form factors. In Section V, we
first present the numerical results of the three 
dynamical quantities as functions of the mass of the strange current
quark and discuss the effects of the flavor $\mathrm{SU}(3)$ symmetry
breaking. Then we show the main results of the three form factors of
the kaon. We also compare the numerical results with the various
experimental data and those from $\chi$PT.  In the final Section we
summarize the results and draw conclusions. 
\section{ Vector and axial-vector form factors of 
the $K^+ \to l^+ \nu   \gamma$ decay }  
The SD part of the kaon radiative decay amplitude is expressed in
terms of the vector and axial-vector form factors of the kaon, i.e.,
$F_V(q^2)$ and $F_A(q^2)$, and the second axial-vector form factor
$R_A(q^2)$. The transition matrix elements of the vector and
axial-vector currents are parametrized by these three form factors
\begin{align}
\langle \gamma(k)|V^{45}_{\mu}(0)|K^+(p)\rangle
&=\frac{e}{m_{K}}\epsilon^{* \alpha} F_{V}(q^2) 
\epsilon_{\mu \alpha \rho \sigma} p^{\rho} k^{\sigma}, 
\label{eq:VectorMarix}
\\
\langle \gamma(k)|A^{45}_{\mu}(0)|K^+(p)\rangle
&=ie\epsilon^{*\alpha}
\sqrt{2}f_K \left[
-g_{\mu \alpha}
+q_{\mu} (q_{\alpha}+p_{\alpha})\frac{ F_{K}(k^2)}{q^2-m_{K}^2}
\right]\cr
&\hspace{0.3cm}
+i\epsilon^{*\alpha}\frac{e }{m_{K}}
\left[
F_A(q^2)\left(k_{\mu}q_{\alpha}-g_{\mu \alpha}q\cdot k\right)
+R_A(q^2)\left(k_{\mu}k_{\alpha}-g_{\mu\alpha}k^2\right)
\right],
\label{eq:AxialMarix}
\end{align}
where $|K^+(p)\rangle$ and $|\gamma (k) \rangle$ denote the
initial kaon and the final photon states. The $\Delta S=1$ transition
vector and axial-vector currents are defined respectively as 
\begin{align}
V^{45}_{\mu} = \bar{\psi} \gamma_\mu \frac{\lambda^4-i \lambda^5}{2}
  \psi,\;\;\;
A^{45}_{\mu} = \bar{\psi} \gamma_\mu\gamma_5 \frac{\lambda^4-i
  \lambda^5}{2}   \psi, 
\label{eq:vacurrent}
\end{align}
where $\psi=(u,\,d,\,s)$ represents the quark field, $\gamma_\mu$ and
$\gamma_5$ the Dirac matrices, and $\lambda^i$ the flavor
$\mathrm{SU}(3)$ Gell-Mann matrices. $p$ and $k$ stand for the
momenta of the kaon and the photon respectively, whereas $q$ is the
momentum of the lepton pair. The value of the kaon mass is taken from
the experimental data $m_K= 493.68$ MeV. The second axial-vector form
factor, $R_A(q^2)$ is considered only when the outgoing photon is
virtual($k^2\ne 0$). $F_{K}(k^2)$ designates the electromagnetic (EM)
form factor which becomes unity at $k=0$, i.e. $F_{K}(0)=1$. Note
that the EM charge radius $\langle r^2_\pi\rangle$ and $F_{K}(k^2)$,
were already computed in this model~\cite{Nam:2007gf}. In the present
work, we concentrate on the derivation of the three form factors
$F_V(q^2)$, $F_A(q^2)$, and $R_A(q^2)$ within the gauged E$\chi$A. 

\section{Gauged nonlocal effective chiral action 
  in the presence of external fields}
\label{sec:3}
Since the EM, vector and axial-vector currents are involved in  
computing the form factors of kaon radiative decay, it is essential to
preserve the relevant gauge invariance. So, we introduce the
corresponding external fields into the E$\chi$A as follows
\begin{align}
\mathcal{S}_{\mathrm{eff}}[v^{\mathrm{em}},v,a,\mathcal{M}]
=-\mathrm{Sp}\ln \left[i\, \slashed{\mathcal{D}}+i\hat{m}
+i\sqrt{M^f(i\,\mathcal{D}^{L})}
  U^{\gamma_5}\sqrt{M^f(i\,\mathcal{D}^{R})} \right], 
\label{eq:GaugeEXA}
\end{align}
where the functional trace $\mathrm{Sp}$ runs over the
space-time, color, flavor, and spin spaces. The current quark mass
matrix $\hat{m}$ is written as $\hat{m}=
\mathrm{diag}(m_{\mathrm{u}},\,m_{\mathrm{d}},\,m_{\mathrm{s}})$. Since 
isospin symmetry is assumed, we set $m_{\mathrm{u}}=m_{\mathrm{d}}$.
The covariant derivative $\mathcal{D}_\mu$ is 
defined as    
\begin{align}
i\,\mathcal{D}_{\mu} =
i\partial_{\mu} + e\hat{Q}\,v^{\mathrm{em}}_{\mu} 
+ \frac{\lambda^a}{2} v^a_{\mu} + \gamma_5 \frac{\lambda^a}{2} a^a_{\mu},
\end{align}
where $v_\mu^{\mathrm{em}}$, $v_\mu^a$, and $a_\mu^a$ denote the
external EM field, vector field and axial-vector field, respectively. 
The charge operator $\hat{Q}$ for the quark fields is defined by
\begin{align}
\hat{Q}= \left(
\begin{array}{ccc}
\frac{2}{3} 	& 0 & 0\\
0        	& -\frac{1}{3} & 0\\
0         & 0 & -\frac{1}{3} 
\end{array}
\right) = \frac{1}{2}\lambda^{3} + \frac{\sqrt{3}}{6}\lambda^{8}.
\end{align}
The left-handed and right-handed covariant derivatives in the
momentum-dependent dynamical quark mass $M(i\mathcal{D}_{L,R})$ are 
defined respectively as 
\begin{align}
i\,\mathcal{D}_{\mu}^L =
i\partial_{\mu} + e\hat{Q}\,v^{\mathrm{em}}_{\mu} 
+ \frac{\lambda^a}{2} v^a_{\mu}  - \gamma_5 \frac{\lambda^a}{2}
  a^a_{\mu},\;\;\;
i\,\mathcal{D}_{\mu}^R =
i\partial_{\mu} + e\hat{Q}\,v^{\mathrm{em}}_{\mu} 
+ \frac{\lambda^a}{2} v^a_{\mu} + \gamma_5 \frac{\lambda^a}{2} a^a_{\mu}.  
\label{eq:covLR}
\end{align}
The covariant derivatives ensure the gauge invariance of
Eq.~(\ref{eq:GaugeEXA}) in the presence of the external
fields~\cite{Musakhanov:2002xa, Kim:2004hd}. The nonlinear
pseudo-Nambu-Goldstone boson field is expressed as
\begin{equation}
\label{eq:NLGB}
U^{\gamma_5} = U(x) \frac{1+\gamma_5}{2} + U^\dagger (x)
\frac{1-\gamma_5}{2} = \exp \left(
i\frac{\gamma_5}{f_\pi}\lambda^\alpha \mathcal{M}^\alpha
\right),
\end{equation}
where $f_\pi$ denotes the pion decay constant and $\mathcal{M}^\alpha$
represent the pseudoscalar meson fields which can be written as 
\begin{equation}
\label{eq:PNGBF}
\lambda^{\alpha} \mathcal{M}^{\alpha} = \sqrt{2} \left( \begin{array}{ccc}
\frac{1}{\sqrt{2}}\pi^0 + \frac{1}{\sqrt{6}}\eta  & \pi^+ & K^+  \\
\pi^- & -\frac{1}{\sqrt{2}}\pi^0  + \frac{1}{\sqrt{6}}\eta & K^0 \\
K^- & \bar{K}^0 & -\frac{2}{\sqrt{6}}\eta 
\end{array}\right).
\end{equation}

The momentum-dependent dynamical quark mass, which arises from the 
quark zero-modes of the Dirac equation in the instanton background fields,
is expressed as
\begin{equation}
\label{eq:DQM}
M^f(k) = M_0 F^2(k) f(m_f),
\end{equation}
where $M_0$ is the constituent quark mass at zero quark virtuality,
and is determined to be around $350$ MeV by the saddle-point
equation~\cite{Diakonov:1985eg,Diakonov:2002fq}. The form factor
$F(k)$ arises from the Fourier transform of the quark 
zero-mode solution for the Dirac equation 
\begin{equation}
\label{eq:ff_dqm}
F(k) = 2\tau\left[ I_0 (\tau) K_1(\tau) -I_1 (\tau)K_0(\tau)
  -\frac{1}{\tau}I_1(\tau)K_1(\tau) \right],
\end{equation}
where $\tau\equiv\frac{|k|\bar{\rho}}{2}$. $I_{i}$ and $K_{i}$
designate the modified Bessel functions. Since it is rather difficult
to use Eq.~\eqref{eq:ff_dqm} for the form factors of the kaon
radiative decays, we employ the dipole-type parametrization of
$F(k)$ defined by  
\begin{equation}
\label{eq:ff_dipole}
F(k)= \frac{2 \Lambda^2}{2 \Lambda^2 + k^2}
\end{equation}
with $\Lambda=1/\bar{\rho}$ together with the modified Bessel
functions. We already haven shown in Ref.~\cite{Shim:2017wcq} that the
results of the form factors for the radiative pion decay are not much
changed by the dipole-type parametrization in place of the original
form. Since $\bar{\rho}=0.33$ fm in the large $N_c$ limit, we take
$\Lambda=600$ MeV. However, we use both Eqs.~\eqref{eq:ff_dqm} and
\eqref{eq:ff_dipole} to obtain the observables apart from the form
factors.   

The presence of the current quark mass
also affects the dynamical one, which was studied in
Refs.~\cite{Musakhanov:1998wp,Musakhanov:2002vu} in detail.  
The additional factor $f(m_f)$ describes the $m_f$ dependence of the
dynamical quark mass, which is defined
as~\cite{Pobylitsa:1989uq,Musakhanov:2001pc}    
\begin{align}
f(m_f) =  \sqrt{1+\frac{m_f^2}{d^2}} -  \frac{m_f}{d}.
\end{align}
This $m_f$-dependent dynamical quark mass produces the gluon
condensate that does not depend on $m_f$~\cite{Nam:2006ng}. Pobylitsa
considered the sum of all planar diagrams, expanding the quark
propagator in the instanton background in the large $N_c$
limit~\cite{Pobylitsa:1989uq}. Taking the limit of $N/(VN_c)\to 0$
leads to $f(m_f)$. The parameter $d$ is given by $198$ MeV. The
$m_f$-dependent dynamical quark mass also explains a correct hierarchy
of the chiral quark condensates: 
$\langle\bar{u}u\rangle\approx\langle\bar{d}d\rangle> 
\langle\bar{s}s\rangle$~\cite{Nam:2006ng}.
The explicit value of the strange current quark mass is actually not a
free parameter. It can be determined by computing the kaonic two-point  
correlation function of the axial-vector currents and fixing the
pole mass by the experimental data. However, the expression for the
kaon mass will be rather complicated, in particular, when the quark
mass is momentum-dependent. Instead, one can also use the
relation~\cite{Gasser:1984gg}  
\begin{align}
\frac{m_K^2}{m_\pi^2} = \frac{m_{\mathrm{s}} + \tilde{m}}{2 \tilde{m}},  
\label{eq:massrel}
\end{align}
where $\tilde{m}$ is the average of the up and down current quark
masses whose values are chosen as $m_u=m_d=5\,\mathrm{MeV}$. Though
Eq.~\eqref{eq:massrel} is valid only at the leading 
order in the large $N_c$ expansion and $m_{\mathrm{s}}$ expansion, we
will adopt it to scrutinize the dependence of the form factors on the
current quark mass for simplicity. The value of the strange current
quark mass is set to be $m_{\mathrm{s}}=120$ MeV~\footnote{In the
  previous works~\cite{Nam:2007fx, Nam:2007gf} within the same
  framework, the mass of the strange current quark is taken to be
  $m_{\mathrm{s}}=150$ MeV as a free parameter.}, which  
satisfies Eq.~\eqref{eq:massrel} very well.
\section{Derivation of the kaon form factors} 
\begin{figure}[htp]
\includegraphics[width=15cm]{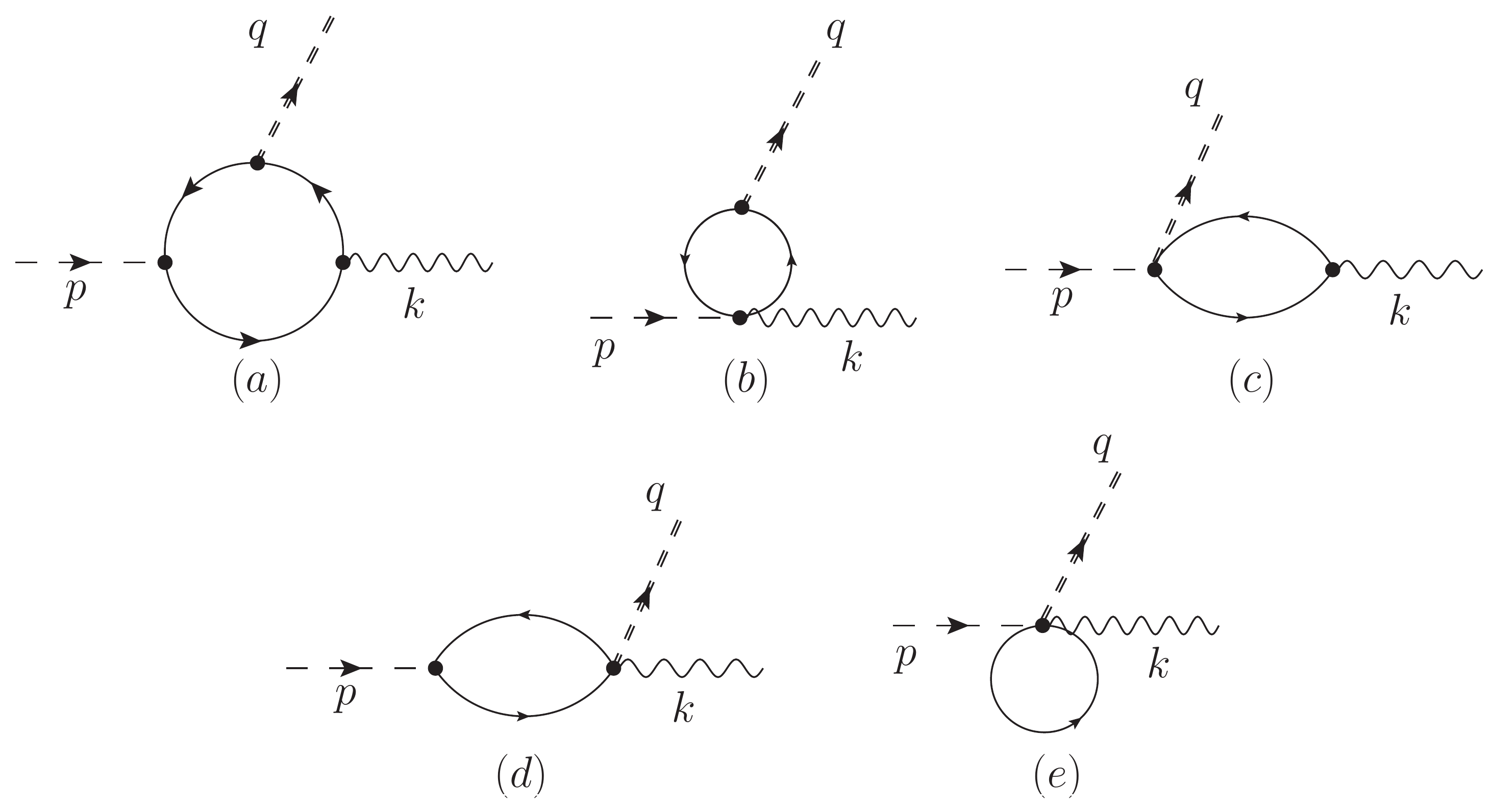}
\caption{The Feynman diagrams for the vector and axial-vector
form factors for the radiative kaon decay. The dashed line depicts the
kaon, the dashed double line and the wavy line describe 
the weak gauge boson (which corresponds to the vector or axial-vector field)
and the photon, respectively. Diagram (a)
contains both the local and nonlocal contributions, whereas diagrams
(b)-(e) arise solely from the nonlocal interaction due to the
momentum-dependent dynamical quark mass.}
\label{fig:1}
\end{figure}

The matrix elements of the vector and axial vector currents given in
Eq.(\ref{eq:AxialMarix}) are derived by computing the following
three-point correlation function 
\begin{align}
\langle\gamma (k)|\mathcal{W}_\mu^{a}(0) |\mathcal{M}^b(p)\rangle
=\epsilon^*_\alpha\int d^4x e^{-ik\cdot x} \int d^4y e^{ip\cdot y}
 \mathcal{G}_{\alpha \rho}^{-1}(k)\mathcal{G}_{\mathcal{M}}^{-1}(p)\langle
  0|\{V^{\mathrm{em}}_{\rho}(x)  \mathcal{W}_{\mu}^{a} (0)
  P^b(y)\}|0\rangle,  
\label{eq:matrixel}
\end{align}
where $\mathcal{W}_\mu^{a}$ denotes generically either the vector
current or the axial-vector current defined in
Eq.(\ref{eq:vacurrent}). The operators $V^{em}_\rho$ and $P^b$ in the
correlation function correspond to the EM current,
and the field operator for the kaon,
respectively. $\mathcal{G}_{\alpha\rho} (k)$,
$\mathcal{G}_{\mathcal{M}} (p)$ stand for the propagators of the
photon and the meson, respectively. The matrix
element~(\ref{eq:matrixel}) can be obtained by differentiating the
gauged E$\chi$A in Eq.(\ref{eq:GaugeEXA}) with respect to the external
EM field, the vector (axial-vector) field, and the kaon field 
\begin{align}
\langle\gamma (k)|\mathcal{W}_\mu^{a}(0) |\mathcal{M}^b(p)\rangle
=\epsilon^*_\alpha\int d^4x e^{-ik\cdot x} \int
  d^4y e^{ip\cdot y}  \left. \frac{\delta^3
  \mathcal{S}_{\mathrm{eff}}[ v^{\mathrm{em}}, w, 
  \mathcal{M} ] }{\delta v_\alpha^{\mathrm{em}}(x) \delta w_\mu^a(0)\delta
  \mathcal{M}^{b}(y)}   \right|_{v^{\mathrm{em}},w,\mathcal{M}=0}, 
\label{eq:corr}
\end{align}
which results in five Feynman diagrams drawn in Fig.~\ref{fig:1}. 
The situation is very similar to the case of the pion. 
Only diagram (a) contributes to the vector form factor, while all
other diagrams vanish because of the trace over spin  
space. On the other hand, all the diagrams come into play when it
comes to the axial-vector form factors.  Diagram (a)
includes both the local and nonlocal terms,
while all other diagrams comes only from the nonlocal terms with the
derivatives of the momentum-dependent quark mass. 

We want to explain the decomposition of the local and nonlocal
parts. As mentioned previously, the presence of the momentum-dependent
dynamical quark mass, which causes a nonlocal interaction, breaks the
U(1) gauge invariance.  Thus, one has to introduce the gauge
connection between two different positions in space-time, which was
done in Ref.~\cite{Musakhanov:2002xa} within the instanton
vacuum. When the strength of the external field is not strong,
Refs.~\cite{Musakhanov:2002xa, Kim:2004hd}  showed that the minimal 
coupling is valid, i.e. the gauge invariance can be restored just by
replacing the ordinary derivative in the action with the covariant
one. Therefore, when one computes the correlation function that
contains gauge fields, one should differentiate the gauged E$\chi$A
not only with respect to the explicit gauge fields in the action but
also with respect to those inside the momentum-dependent dynamical
quark mass. When we turn off the momentum dependence of the dynamical
quark mass, the contributions, which arise from the derivative of
$M^f(iD)$ with respect to the gauge field, vanish. We call these
contributions as nonlocal ones. Those coming from the gauge fields out
of $M^f$ are called the local contributions. 
\subsection{Vector form factor}
We first derive the vector form factor of the kaon. Having computed  
Eq.(\ref{eq:corr}) explicitly, we obtain the matrix element of the
vector current ($\mathcal{W}=V$)  
\begin{align}
\langle\gamma(k)|V_{\mu}^{45}|K^+(p)\rangle
&= i\frac{4\sqrt{2}e N_c}{3f_\pi}
\epsilon^{*}_{\alpha} \int
\frac{d^4 l}{(2\pi)^{4}}
\frac{
\sqrt{M^u(k_a) M^s(k_b)}
}{
\mathcal{D}^u_{a}\mathcal{D}^s_{b}\mathcal{D}^s_{c}
}
\left[\varepsilon_{\mu\alpha\rho\sigma}
\left(\bar{M}^u_a k_{b\rho} k_{c\sigma} + \bar{M}^s_b k_{c\rho} k_{a\sigma}
+ \bar{M}^s_c k_{a\rho} k_{b\sigma} \right) \right.
\cr
&\hspace{0.4cm} 
+ \varepsilon_{\alpha\beta\rho\sigma} k_{a\beta} k_{b\rho} k_{c\sigma} 
\left(\sqrt{M^u(k_a)}\sqrt{M^u_{\mu}(k_a)} 
+ \sqrt{M^s(k_c)}\sqrt{M^s_{\mu}(k_c)} \right)
\cr
&\hspace{0.4cm}
\left.- \varepsilon_{\mu\beta\rho\sigma} k_{a\beta} k_{b\rho} k_{c\sigma}
\left( \sqrt{M^s(k_b)}\sqrt{M^s_{\alpha}(k_b)} 
+\sqrt{M^s(k_c)}\sqrt{M^s_{\alpha}(k_c)} \right)
\right] -2(u\leftrightarrow s),
\label{eq:VecMatElmt}
\end{align}
where $N_c$ is the number of colors and $\bar{M}^f_i$ is defined by
the sum of the dynamical and current quark masses
$\bar{M}^f_i=m_f+M^f(k_i)$. The momenta $k_i$ are expressed as 
$k_a=l+\frac{q}{2}+\frac{k}{2}$, $k_b=l-\frac{q}{2}-\frac{k}{2}$,
$k_c=l-\frac{q}{2}+\frac{k}{2}$, and $q=p-k$. 
$\mathcal{D}^f_i$ are given as $\mathcal{D}^f_{i}=(k_i^2+\bar{M}^{f\,
  2}_i)$. $\sqrt{M^f_\mu(k_i)}$ represents $\sqrt{M^f_\mu(k_i)}
= \partial \sqrt{M^f(k_i)}/\partial
k_{i\mu}$. Equation~(\ref{eq:VecMatElmt}) corresponds to diagram (a) 
in Fig.~\ref{fig:1} and diagrams (b)-(e) do not contribute at all to
the vector form factor. 
Using the transverse condition $\epsilon^{*}\cdot p=\epsilon
\cdot p=0$, we are able to extract the vector form factor, comparing
Eq.(\ref{eq:VectorMarix}) with Eq.(\ref{eq:VecMatElmt}). 
The final expression of the vector form factor is written as 
\begin{align}
F_V(Q^2) = F_V^{\mathrm{local}} (Q^2) + F_V^{\mathrm{NL}}(Q^2),   
\end{align}
where $F_V^{\mathrm{local}} (Q^2)$ and $F_V^{\mathrm{NL}}(Q^2)$
stand for the local and nonlocal contributions respectively
\begin{align}
F_V^{\mathrm{local}}(Q^2) = \frac{4\sqrt{2} M_{K}}{f_\pi}
                            \mathcal{G}_V^{\mathrm{local}}(Q^2),\;\;\;\;
F_V^{\mathrm{NL}}(Q^2) = \frac{4\sqrt{2} M_{K}}{f_\pi}
                            \mathcal{G}_V^{\mathrm{NL}}(Q^2).
\end{align}
Here, $\mathcal{G}_V^{\mathrm{local}}$ and
$\mathcal{G}_V^{\mathrm{NL}}$ are defined respectively by
\begin{align}
\mathcal{G}_V^{\mathrm{local}}(Q^2) &= -\frac{N_c}{3(p\cdot k)^2}
\int \frac{d^4 l}{(2\pi)^{4}} 
\frac{\sqrt{M^u(k_a)M^s(k_b)}}{D^u_{a} D^s_{b} D^s_{c}} p_{\mu}
  k_{\nu}\bigg[ \bar{M}^u_a ( k_{b\mu} k_{c\nu} - k_{c\mu} k_{b\nu} )  
\cr
& \hspace{4.4cm}+ \bar{M}^s_b ( k_{c\mu} k_{a\nu} - k_{a\mu} k_{c\nu} ) 
+ \bar{M}^s_c ( k_{a\mu} k_{b\nu} - k_{b\mu} k_{a\nu} )  
\bigg]-2(u\leftrightarrow s),
\label{eq:localv}
\\
\mathcal{G}_V^{\mathrm{NL}} (Q^2)
&=-\frac{ N_c }{3 (p\cdot k)^2}
\int\frac{d^4 l}{(2\pi)^{4}}
\frac{\sqrt{M^u(k_a)M^s(k_b)}}{\mathcal{D}^u_{a}
  \mathcal{D}^s_{b}\mathcal{D}^s_{c}}   
\bigg[-\Big({M^s}'_b+{M^s}'_c\Big)(p\cdot k)^2
(\epsilon^{*}\cdot l)(\epsilon\cdot l) 
\cr
&\hspace{4.4cm}
+\Big( {M^u}'_a + {M^s}'_c \Big) 
(\varepsilon_{\mu\gamma\delta\lambda}l_{\mu}\epsilon_{\gamma}
p_{\delta}k_{\lambda})
(\varepsilon_{\alpha\beta\rho\sigma}\epsilon^{*}_{\alpha}
k_{a\beta}k_{b\rho}k_{c\sigma})
\bigg]-2(u\leftrightarrow s),
\label{eq:nonlocalv}
\end{align}
where ${M^f}'_i$ denotes the derivative of the dynamical quark mass with
respect to the squared momentum, ${M^f}'_i=\partial M^f(k_i)/ \partial
k_i^2 $. We use the positive definite $Q^2$ defined as $Q^2=-q^2$. 
In fact, $\mathcal{G}_V^{\mathrm{local}}$ and
$\mathcal{G}_V^{\mathrm{NL}}$ are exactly same as
those for the pion form factors given explicitly in
Ref.~\cite{Shim:2017wcq} except for the current mass of the strange  
quark. Thus, the results of $\mathcal{G}_V^{\mathrm{local}}$ and
$\mathcal{G}_V^{\mathrm{NL}}$ will exhibit how the vector form factor
gets changed, if one considers the mass of the strange current quark
as a parameter. 

Note that the terms with $M'(k_i)$ are derived from the expansion
of the dynamical quark mass with respect to the covariant 
derivative given in Eq.~(\ref{eq:covLR}). Thus, those terms with
$M'(k_i)$ are the essential part in obtaining the vector and
axial-vector form factors with the corresponding gauge invariance
preserved, which are called the nonlocal contributions as explained
previously already. If the dynamical quark mass is considered to be
constant, then $M'(k_i)$ vanishes. However, one has to keep in mind
that certain regularizations must be introduced to tame the divergence
arising from the quark loop in the local $\chi$QM. Hence, the
momentum-dependent dynamical quark mass can be also regarded as a
regularization. 

\subsection{Axial-vector form factors}
Similarly, the transition matrix element of the axial-vector current
 ($\mathcal{W}=A$) in Eq.(\ref{eq:AxialMarix}) can be obtained as  
\begin{align}
\langle\gamma(k)|A_{\mu}^{45}|K^+(p)\rangle
=-i\frac{4\sqrt{2}eN_c}{3f_{\pi}}
\epsilon^{*}_{\alpha}\int\frac{d^4 l}{(2\pi)^4}
\sum^{e}_{i=a}
\left[\left(\mathcal{F}^{(i)}_{\mu\alpha} + 2(u\leftrightarrow s)\right)
+\left(\tilde{\mathcal{F}}^{(i)}_{\mu\alpha} - (u\leftrightarrow s)\right)
\right],
\label{eq:axialVFF1}
\end{align}
where $\mathcal{F}^{(i)}_{\mu\alpha}$ and
$\tilde{\mathcal{F}}^{(i)}_{\mu\alpha}$ correspond to diagram
($i$) in which $i=a,\,b,\,c,\,d,\,e$. Explicitly, they can be written as
\begin{align}
\mathcal{F}^{(a)}_{\mu\alpha}
&=\frac{\sqrt{M^u(k_a)M^s(k_b)}}{\mathcal{D}^u_{a}\mathcal{D}^s_{b}
  \mathcal{D}^s_{c}} 
\left[\delta_{\mu\alpha}  \left\{ 
\bar{M}^u_a k_{b}\cdot k_{c} 
-\bar{M}^s_b k_{c}\cdot k_{a} +\bar{M}^{s}_c k_{a}\cdot k_{b}+\bar{M}^{uss}_{abc}
\right\}\right. \cr
&+\left\{
-\bar{M}^u_a(k_{b\mu}k_{c\alpha} + k_{c\mu}k_{b\alpha})
+\bar{M}^s_b(k_{a\mu}k_{c\alpha}+k_{c\mu}k_{a\alpha})
+\bar{M}^s_c(k_{a\mu}k_{b\alpha}-k_{b\mu}k_{a\alpha})
\right\}\cr
& + \left(
{M^s}'_b k_{b\alpha}+{M^s}'_c k_{c\alpha}\right)\left\{
-(k_{b}\cdot k_{c}-\bar{M}^{ss}_{bc})k_{a\mu}
+(k_{c}\cdot k_{a}-\bar{M}^{su}_{ca})k_{b\mu}
-(k_{a}\cdot k_{b}+\bar{M}^{us}_{ab})k_{c\mu}\right\} \cr
&-\left(
{M^u}'_a k_{a\mu}-{M^s}'_c k_{c\mu}\right)\left\{
 (k_{b}\cdot k_{c}+\bar{M}^{ss}_{bc})k_{a\alpha}
-(k_{c}\cdot k_{a}+\bar{M}^{su}_{ca})k_{b\alpha} 
-(k_{a}\cdot k_{b}+\bar{M}^{us}_{ab})k_{c\alpha}\right\}\cr
& -\left. \left({M^s}'_b k_{b\alpha}+{M^s}'_c k_{c\alpha}\right)
\left({M^u}'_a k_{a\mu}-{M^s}'_c k_{c\mu}\right)\left\{
\bar{M}^u_a k_{b}\cdot k_{c}
-\bar{M}^s_b k_{c}\cdot k_{a}
-\bar{M}^s_c k_{a}\cdot k_{b}-\bar{M}^{uss}_{abc}
\right\}\right], \cr
\mathcal{F}^{(b)}_{\mu\alpha}
&=\frac{1}{\mathcal{D}^u_{a}\mathcal{D}^s_{c}}
\sqrt{M^u(k_a)} \sqrt{M^s_{\alpha}(k_b)}
\left[-\left\{\bar{M}^s_c+{M^u}'_a (k_{a}\cdot k_{c}+\bar{M}^{us}_{ac})\right\}
  k_{a\mu}
+\left\{\bar{M}^u_a+{M^s}'_c(k_{a}\cdot
  k_{c}+\bar{M}^{us}_{ac})\right\} k_{c\mu}\right], \cr 
\mathcal{F}^{(c)}_{\mu\alpha}
&=\frac{1}{\mathcal{D}^{s}_{b}\mathcal{D}^{s}_{c}}
\sqrt{M^u_{\mu}(k_a)}\sqrt{M^s(k_b)}\left[
-\left\{\bar{M}^s_c-{M^s}'_b (k_{b}\cdot k_{c}-\bar{M}^{ss}_{bc})\right\}
  k_{b\alpha} 
-\left\{\bar{M}^s_b-{M^s}' _c (k_{b}\cdot k_{c}-\bar{M}^{ss}_{bc})\right\}
  k_{c\alpha} \right], \cr
\mathcal{F}^{(d)}_{\mu\alpha}
&=\frac{1}{\mathcal{D}^u_{a}\mathcal{D}^s_{b}}
\sqrt{M^u(k_a)M^s(k_b)}\sqrt{M^s_{\mu}(k_c)}\sqrt{M^s_{\alpha}(k_c)}
\left(k_{a} \cdot k_{b} +\bar{M}^{us}_{ab}\right), \cr
\mathcal{F}^{(e)}_{\mu\alpha}
&=\frac{\bar{M}^s_c}{\mathcal{D}^s_c}
\sqrt{M^u_\mu(k_a)}\sqrt{M^s_\alpha(k_b)}
\label{eq:fas}
\end{align}
and
\begin{align}
&\tilde{\mathcal{F}}^{(a)-(c)}_{\mu\alpha}
=0, \cr
&\tilde{\mathcal{F}}^{(d)}_{\mu\alpha}
=-\frac{M^s(k_b)}{2\mathcal{D}^u_a\mathcal{D}^s_b}
\sqrt{M^u(k_a)}\sqrt{M^s_{\mu\alpha}(k_b)}
\left(k_{a} \cdot k_{b} +\bar{M}^{us}_{ab}\right), \cr
&\tilde{\mathcal{F}}^{(e)}_{\mu\alpha}
=-\frac{\bar{M}^s_b}{2\mathcal{D}^s_b}\sqrt{M^u_{\mu\alpha}(k_a)}\sqrt{M^s(k_b)}.
\label{eq:fSBas}
\end{align}
Here, we have introduced the following short-handed notations
$\bar{M}^{f_1 f_2}_{ij}=\bar{M}^{f_1}_i \bar{M}^{f_2}_j$, 
$\bar{M}^{f_1 f_2 f_3}_{ijk}= \bar{M}^{f_1}_i
\bar{M}^{f_2}_j \bar{M}^{f_3}_k$ 
and
$\sqrt{M^f_{\mu\alpha}(k_i)}=\partial^2\sqrt{M^f(k_i)}/\partial
k_{i\mu} \partial k_{i\alpha}$. 

As done in the case of the pion axial-vector form factors,  the kaon 
axial-vector form factors can be easily obtained by introducing an 
arbitrary vector $\xi^\perp_{\mu}$ that satisfies the following
properties: $\xi^\perp \cdot \xi^\perp =0$, $\xi^\perp \cdot q=0$, and
$\xi^\perp \cdot k\ne0$. Hence, the axial-vector form
factors $F_A(Q^2)$ and $R_A(Q^2)$ are obtained as 
\begin{align}
F_{A}(Q^2)
=\frac{4\sqrt{2} m_K}{ f_\pi }
\mathcal{G}_A(Q^2)
, \quad
R_{A}(Q^2)
=\frac{4\sqrt{2} m_K}{f_\pi}
\mathcal{H}_A(Q^2)
\end{align}
where
\begin{align}
\mathcal{G}_{A}(Q^2)
&=\frac{N_c}{3 (q \cdot k)}
\int \frac{d^4 l}{(2 \pi)^4}
\sum^{e}_{i=a}\left[\left(\mathcal{F}^{(i)}_{\mu\alpha} + 2(u\leftrightarrow s)\right)
+\left(\tilde{\mathcal{F}}^{(i)}_{\mu\alpha} - (u\leftrightarrow s)\right)
\right]
\left( 
\frac{\xi^\perp_{\mu} k_{\alpha}}{\xi^\perp \cdot k}
-\epsilon_\mu \epsilon^{*}_{\alpha} 
\label{eq:ga}
\right),
\\
\mathcal{H}_{A}(Q^2)
&=\frac{N_c}{3 (\xi^\perp \cdot k)^2}
\int \frac{d^4 l}{(2 \pi)^4}
\sum^{e}_{i=a}\left[\left(\mathcal{F}^{(i)}_{\mu\alpha} + 2(u\leftrightarrow s)\right)
+\left(\tilde{\mathcal{F}}^{(i)}_{\mu\alpha} - (u\leftrightarrow s)\right)
\right]
\xi^\perp_\mu \xi^\perp_\alpha.
\label{eq:ha}
\end{align}
The local contribution to the axial-vector form factors comes from the
first and second terms of $\mathcal{F}_{\mu \alpha}^{(a)}$ in
Eq.~(\ref{eq:fas}).  In the flavor SU(3) symmetric case, that is,
$m_{\mathrm{u}} = m_{\mathrm{d}} = m_{\mathrm{s}}$,
$\tilde{\mathcal{F}}_{\mu \alpha}^{(d,e)}$ would have been exactly
canceled by the exchange terms expressed as $(u\leftrightarrow
s)$. This will lead to the same expressions of the form factors for
the radiative pion decay~\cite{Shim:2017wcq}. Thus,
$\tilde{\mathcal{F}}_{\mu \alpha}^{(d,e)}$ become finite only in the
case of radiative kaon decays. 

\section{Results and discussion}
Before we discuss the results, we
want to emphasize that the gauged E$\chi$A does not contain any
additional parameters to fit. The average size of the instanton
$\rho=\Lambda^{-1}$ and the interdistance between instantons $R$ were
taken from the original work~\cite{Diakonov:1985eg}. The dynamical
quark mass at the zero quark quark virtuality was determined by the
saddle-point equation~\cite{Diakonov:1985eg}. The strange current
quark mass was also fixed by the leading-order mass
relation in the large $N_c$ limit and in the current quark mass
expansion. Nevertheless, a possible uncertainty may arise from the
fact that we employ the dipole-type parametrization of $F(k)$ 
defined in Eq.~\eqref{eq:ff_dipole} instead of the original form
derived from the instanton vacuum. However, as discussed already in
Ref.~\cite{Shim:2017wcq}, the uncertainty for $F_V^\pi$ 
is below $4~\%$, whereas those for $F_A^\pi$ and $R_A^\pi$ are around
$10~\%$  and below $1~\%$, respectively. Thus, the use of the
dipole-type form factor does not affect the conclusion of the present
work.  Otherwise, we do not have any room for changing the
parameters. Note that we do not include the $1/N_c$ meson-loop
corrections~\cite{Kim:2005jc, Goeke:2007bj} in the present work. 

We begin by examining the effects of the flavor $\mathrm{SU}(3)$
symmetry breaking.  The expressions of both the pion and kaon form
factors contain 
the prefactors $m_\pi/f_\pi\approx 1.5$ and $m_K/f_\pi \approx 5.3$
respectively, which makes a large difference in the magnitudes of the
pion and kaon form factors. If we factor out these kinematical factors
and release the value of the strange current quark mass from
$m_{\mathrm{s}}=120$ MeV, then we can more closely explore the effects
of the flavor SU(3) symmetry breaking. Thus, we first compute
$\mathcal{G}_V$, $\mathcal{G}_A$, and $\mathcal{H}_A$ defined in
Eqs.~\eqref{eq:localv}, \eqref{eq:nonlocalv}, \eqref{eq:ga}, and
\eqref{eq:ha}, respectively. 

\begin{figure}[htp]
\captionsetup[subfigure]{labelformat=empty}
\subfloat[]{\includegraphics[width = 2.4in]{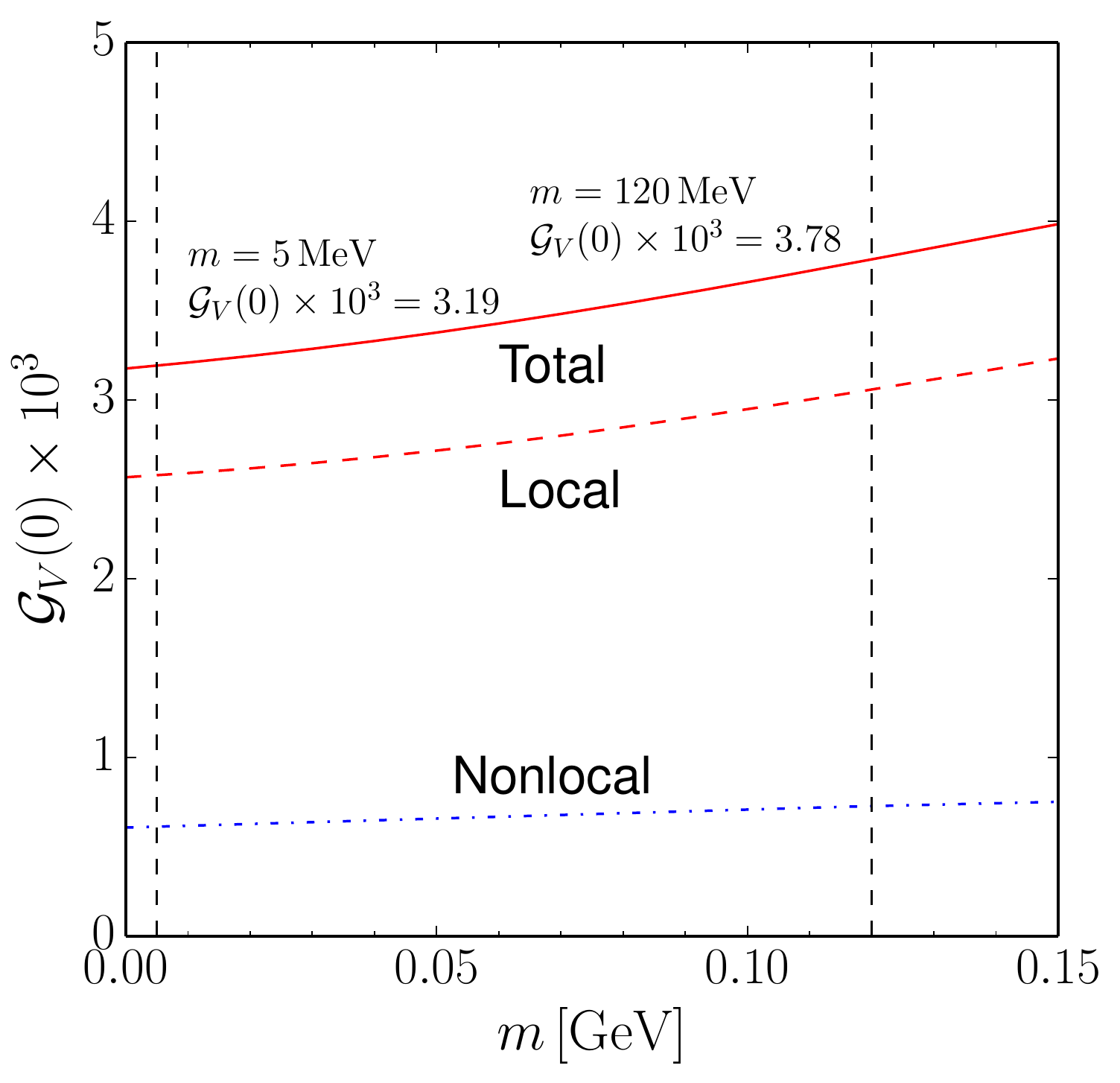}} 
\subfloat[]{\includegraphics[width = 2.4in]{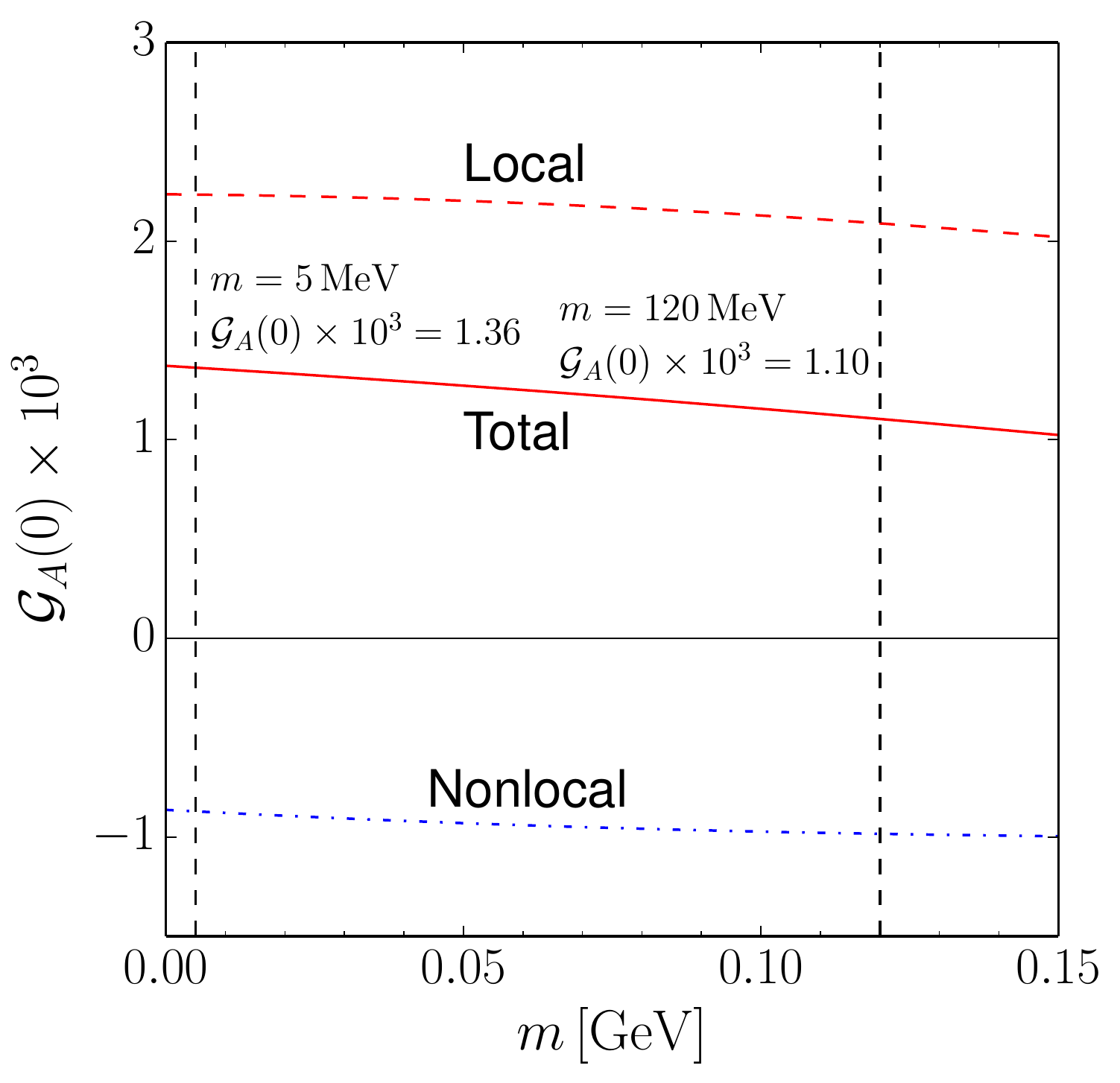}}
\subfloat[]{\includegraphics[width = 2.4in]{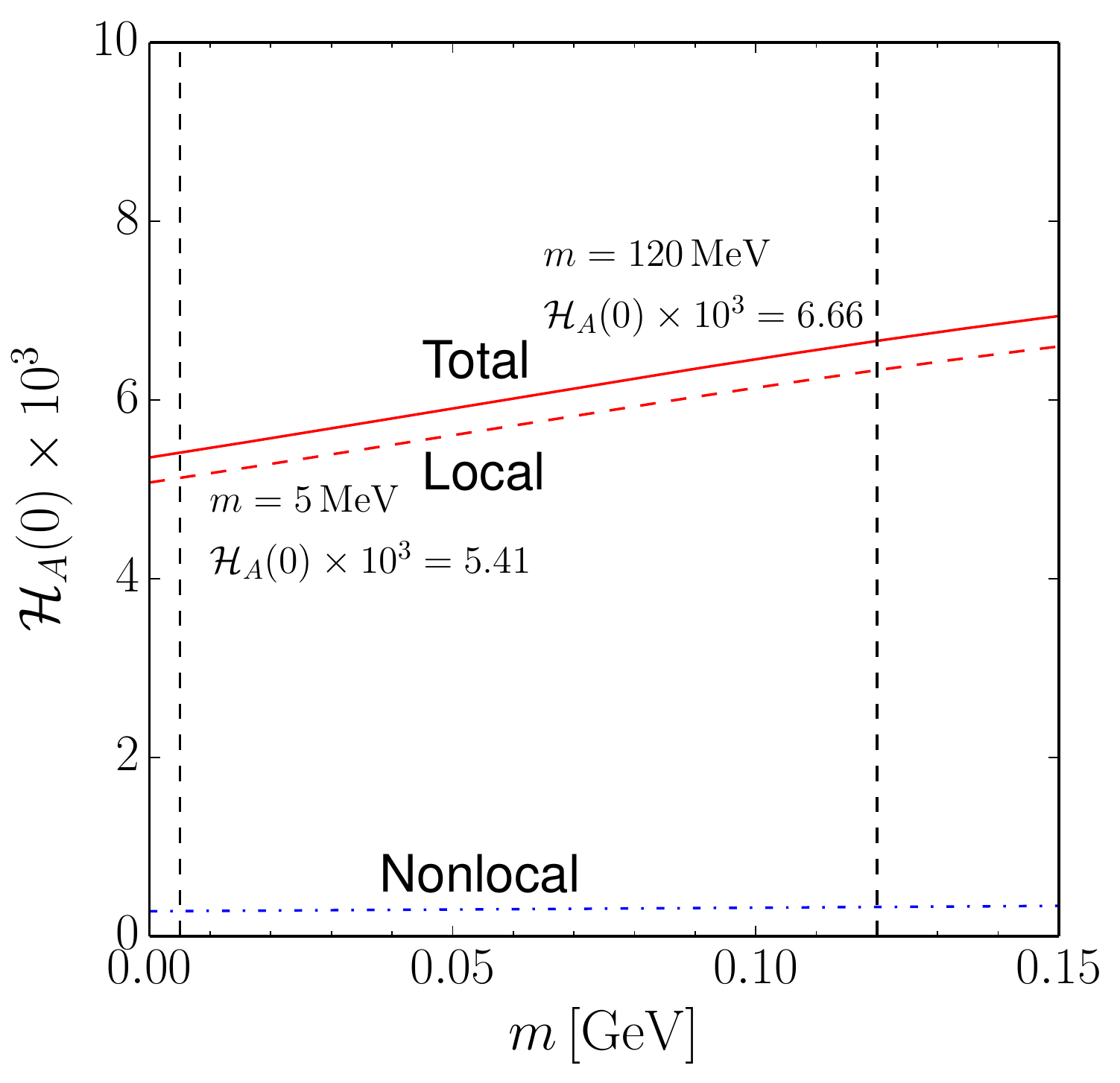}}
\caption{Numerical results of $\mathcal{G}_V(0)$, $\mathcal{G}_A(0)$,
  and $\mathcal{H}_A(0)$ multiplied by $10^3$ as functions of the
  current quark mass $m$, which are drawn respectively in the left,
  middle, and right panels. The dotted and dot-dashed lines depict the 
  local and nonlocal contributions, respectively. The solid line
  represents the total contribution.}
\label{fig:2}
\end{figure}
Figure~\ref{fig:2} draws the numerical results of
$\mathcal{G}_V(0)$, $\mathcal{G}_A(0)$, and $\mathcal{H}_A(0)$ as
functions of a current quark mass $m$ in the range between  $m=5$ MeV
and $m=120$ MeV, in which the effects of flavor SU(3)
symmetry breaking are clearly exhibited. 
In the case of $\mathcal{G}_V$, the local contribution increases
monotonically as the value of $m$ grows. The magnitude of the 
nonlocal part gets only slightly larger when $m$ increases. Hence, the
$m$ dependence is governed by the local terms.
$\mathcal{H}_A$ depicted in the right panel of Fig.~\ref{fig:2} shows
a similar tendency to $\mathcal{G}_V$. On the other hand,
$\mathcal{G}_A$ decreases as $m$ increases. As displayed in
Fig.~\ref{fig:2}, the effects of the SU(3) symmetry breaking
increase the vector form factor $F_V(0)$ and the second
axial-vector form factor $R_A(0)$ for the 
radiative kaon decay by about $16~\%$ and $19~\%$, compared to the 
corresponding pion form factors. On the other hand, they lessen the
axial-vector form factor $F_A(0)$ for the radiative kaon decay in
comparison with the corresponding pion axial-vector form factor.
These kinds of tendencies of $\mathcal{G}_V(0)$, $\mathcal{G}_A(0)$,
and $\mathcal{H}_A(0)$ determine the pion and kaon weak form factors
at $Q^2=0$ with corresponding current quark masses, $m=5$ MeV and
$m=120$ MeV. 

\begin{table}[htp]
\centering
\caption {Results of the form factors at $Q^2=0$ and slope parameters
  in comparison with those from $\chi$PT to order $\mathcal{O}(p^6)$
  and the experimental data. The results listed in the column denoted
  by ``D.P.'' are obtained by using the original momentum-dependent quark
  mass \eqref{eq:ff_dqm} derived in Ref.~\cite{Diakonov:1985eg},
  whereas those in the last column designated by ``Dipole'' are
  obtained by using the dipole-type form factor given in
  Eq.~\eqref{eq:ff_dipole}.}  
\label{tab:1} 
\begin{tabular}{ccm{3.5cm}m{2.5cm}m{4cm}|cc}\hline\hline 
& \multirow{2}{*}{CHPT$p^6$~\cite{Geng:2003mt}} 
& \multicolumn{3}{c|}{$\hspace{-0.8cm}$Experimental data} 
& \multicolumn{2}{c|}{Present results}\\\cline{3-7}  & & 
 $K\to e(\mu) \nu e^+e^-$~\cite{Poblaguev:2002ug} &
 $K\to e\nu\gamma$ & $K\to \mu\nu\gamma$ & D.P. & Dipole \\ \hline
 $F_{V}(0)$        & 0.078(5) & 0.112(28) &   &        & 0.118 & 0.114 \\
 $F_{A}(0)$        & 0.034    & 0.035(30) &   &        & 0.027 & 0.033  \\
 $R_A(0)$          &          & 0.227(32) &   &        & 0.201 & 0.200\\
 $F_V(0)+F_A(0)$ & 0.112(5) & 0.147(40) &
 0.125(8)~\cite{Ambrosino:2009aa}  & 0.165(18)~\cite{Adler:2000vk} &
 0.145 & 0.147 \\    
 $F_V(0)-F_A(0)$ & 0.044(5) & 0.077(45) &   & 0.21(8)\cite{Duk:2010bs}, 
0.126(74)\cite{Tchikilev:2010wy}    &  0.092 & 0.081 \\
 $R_A(0)+F_V(0)$ &          & 0.338(45) &     &        & 0.319 & 0.314 \\
 $R_A(0)-F_V(0)$ &          & 0.114(42) &     &        & 0.083 & 0.086 \\
 $R_A(0)+F_A(0)$ &          & 0.262(21) &     &        & 0.228 & 0.233 \\
 $R_A(0)-F_A(0)$ &          & 0.191(61) &     &        & 0.174 & 0.167 \\
 $a_{V}$         &  0.3(1)  &
&0.38(4)~\cite{Ambrosino:2009aa}  
&    & 0.404 & 0.379 \\
 $a_{A}$         &          &           &                                 
&    & 0.159 & 0.192 \\
  \hline \hline
 \end{tabular}
\end{table}

In Table~\ref{tab:1}, we list the results of the form factors at
$Q^2=0$, various combinations of them, and slope parameters $a_V$ and
$a_A$ in comparison with those from $\chi$PT to order
$\mathcal{O}(p^6)$ and the experimental data. The slope parameters are
defined from the following parametrizations of the vector and
axial-vector form factors for the radiative kaon decay 
\begin{align}
F_{V}(Q^2)=\frac{F_{V}(0)}{1+ a_{V} \frac{Q^2}{m_{K}^2}},\;\;\;
F_{A}(Q^2)=\frac{F_{A}(0)}{1+ a_{A} \frac{Q^2}{m_{K}^2}},
\label{eq:SingPolFit}
\end{align} 
where $a_{V}$ and $a_A$ denote the slope 
parameters for the vector and axial-vector form factors, respectively.
The results listed in the column denoted by ``D.P.'' are obtained by
using the original momentum-dependent quark 
mass defined in Eq.~\eqref{eq:ff_dqm}~\cite{Diakonov:1985eg},
whereas those in the last column designated by ``Dipole'' are 
produced by employing the dipole-type form factor given in
Eq.~\eqref{eq:ff_dipole}. The results with the two different form
factors are not much different from each other. 

Generally, the present numerical results are in very good agreement
with the experimental data taken from Ref.~\cite{Poblaguev:2002ug},
where kaon radiative decays $K^+\to \mu^+ \nu e^+ e^-$ and $K^+\to
e^+\nu e^+ e^-$ were experimentally studied. The experimental data
presented in the third column of Table~\ref{tab:1} are those from the
combined fit including both radiative decays $K^+\to \mu^+ \nu e^+
e^-$ and $K^+\to e^+\nu e^+
e^-$~\cite{Poblaguev:2002ug}. Experimentally, more plausible
quantities are $F_V(0)+F_A(0)$ and $F_V(0)-F_A(0)$. The experimental
data indicate consistently that the decay $K\to \mu \nu
\gamma$ yields the larger values of $F_V+F_A$ than the electron channel
$K\to e \nu \gamma$. For example, Ref.~\cite{Poblaguev:2002ug}
reported the mean value of $F_V+F_A=0.155$ from the $K^+\to \mu^+ \nu 
e^+ e^-$ data whereas $F_V+F_A=0.125$ from the $K^+\to e^+ \nu
e^+ e^-$ data and the weighted average value is in
TABLE~\ref{tab:1}. The results of $F_V-F_A$ show similar tendencies.  
The comparison of the KLOE data~\cite{Ambrosino:2009aa} with those of
the E787 Experiment~\cite{Adler:2000vk} leads to the same
conclusion. The data on $K^-\to \mu^-\nu \gamma$ from the ISTRA+
Collaboration~\cite{Duk:2010bs} gives a rather large value of
$F_V-F_A$, i.e. $F_V-F_A=0.21$ that is almost three times larger than
that from Ref.~\cite{Poblaguev:2002ug}. Another analysis from the 
ISTRA+ Collaboration yields $F_V-F_A=0.126$ with the exotic tensor
interaction excluded~\cite{Tchikilev:2010wy}. The present results lie
between those from the $\mu$ and electron channels. In general, the
results from $\mathcal{O}(p^6)$ $\chi$PT are underestimated, compared
with the present ones except for $F_A$ of which the value is almost
the same as our result.  
The vector slope parameter $a_V$ is experimentally known to be
$a_V=0.38(4)$ from Ref.~\cite{Ambrosino:2009aa} where as $a_A$ is
still at large experimentally. The present results of $a_V$ are
$0.404$ and $0.379$, which are in good agreement with the KLOE data. We
predict $a_A=0.159$ (D.P.) and $a_A=0.192$ (Dipole).  

\begin{figure}[htp]
\captionsetup[subfigure]{labelformat=empty}
\subfloat[]{\includegraphics[width = 2.4in]{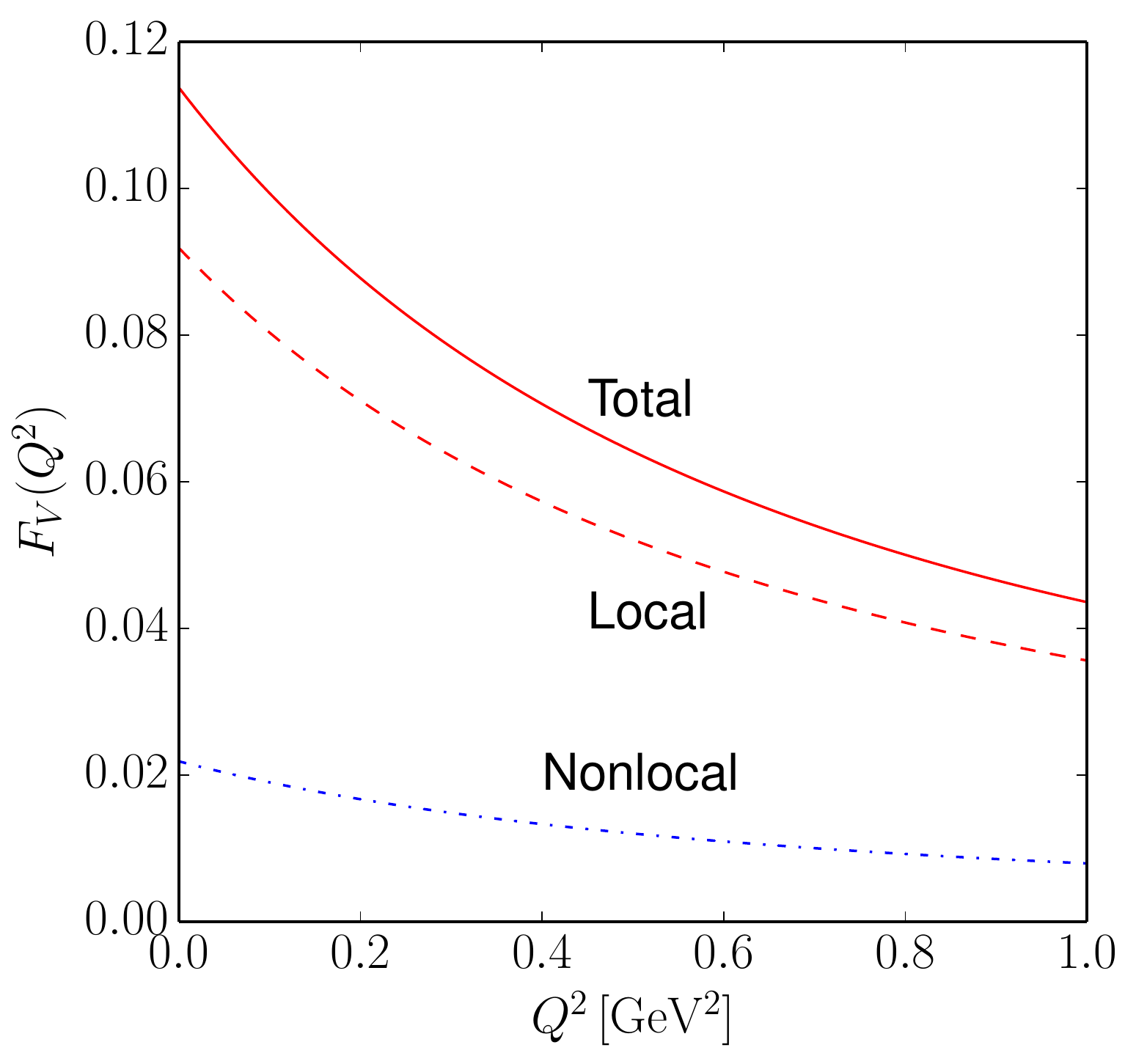}} 
\subfloat[]{\includegraphics[width = 2.4in]{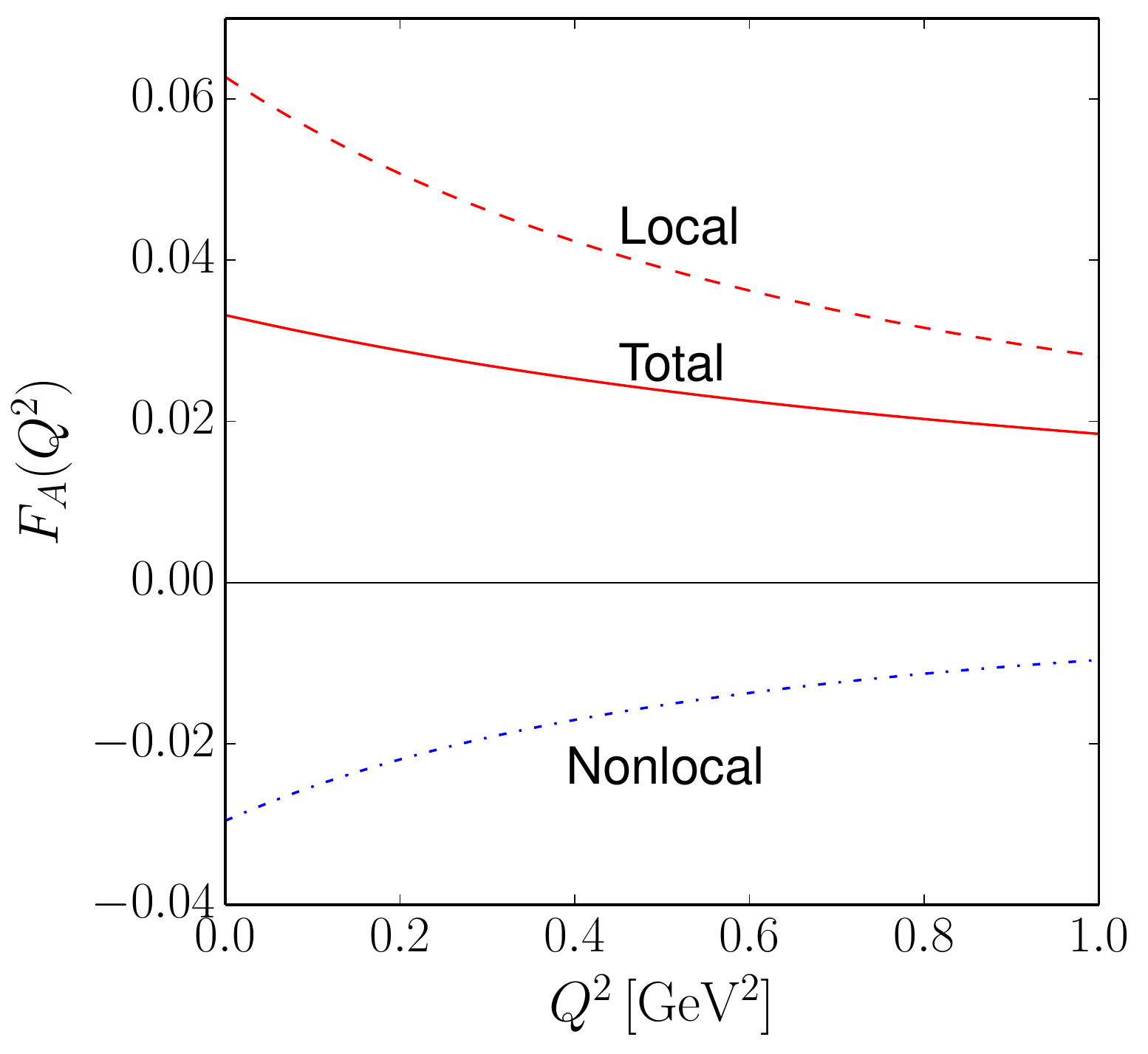}}
\subfloat[]{\includegraphics[width = 2.4in]{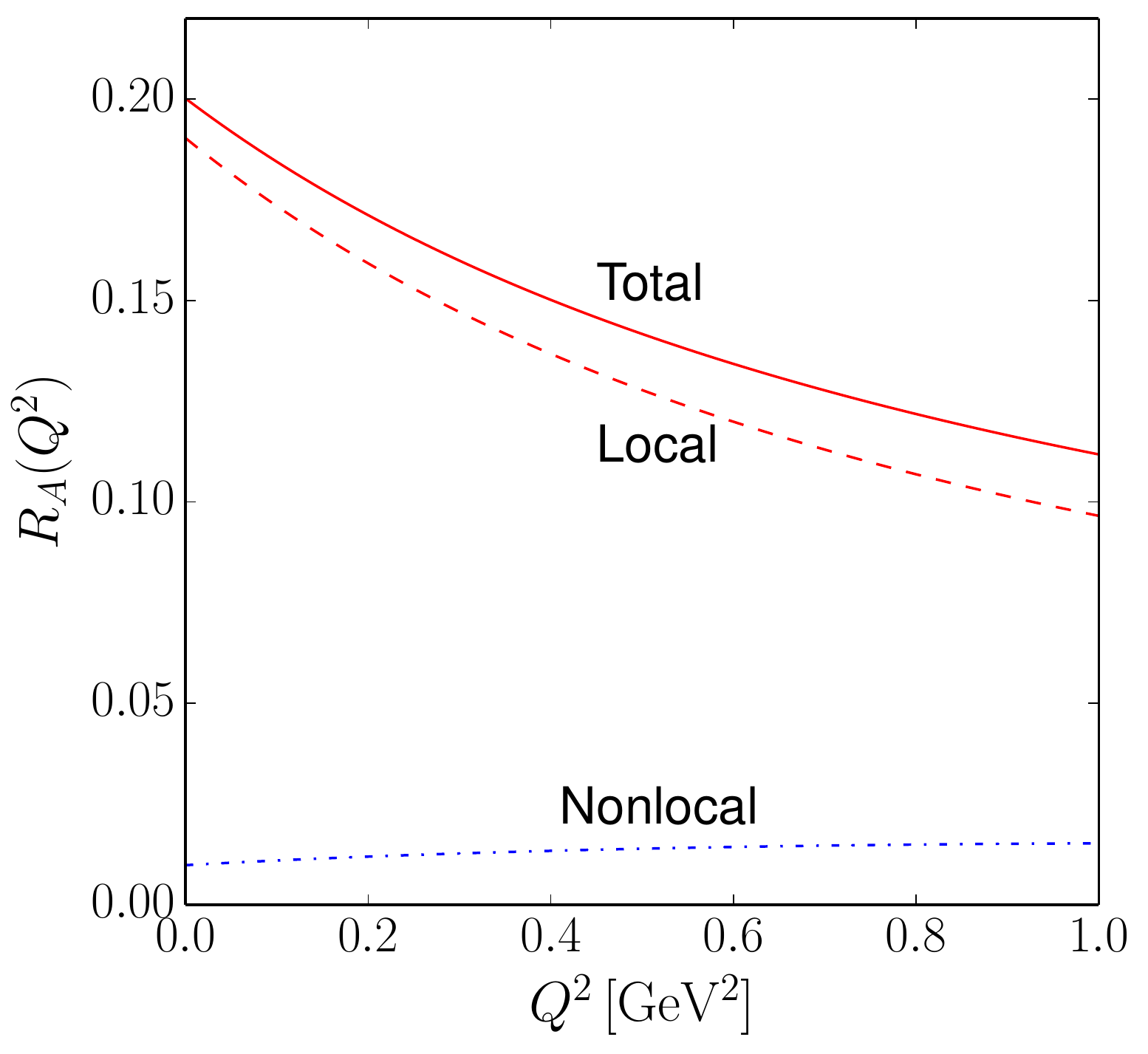}}
\caption{Numerical results of the vector and axial-vector form factors
  for the radiative kaon decays as functions of $Q^2$. The dashed and
  dot-dashed curves depict respectively the local and nonlocal
  contributions, whereas the solid draws the total result.  
  }
\label{fig:3}
\end{figure}

In Fig.~\ref{fig:3} we show the numerical results of the vector and
axial-vector form factors of the kaon radiative decays. All these
three form factors fall off monotonically as $Q^2$ increases. In fact,
the results of the form factors for the radiative kaon decays show the
same tendency as those for the radiative pion
decays~\cite{Shim:2017wcq}, since the expressions of the form factors
are the same as those for the pion decays except for the strange
current quark mass, as already discussed in Fig.~\ref{fig:2}.
Nevertheless let us recapitulate briefly what we have found.  
The nonlocal terms appear from the \emph{gauged} E$\chi$A that was
constructed in such a way that the relevant gauge invariance is
preserved. In the left panel of Fig.~\ref{fig:3}, we find that the
nonlocal contribution enhances the vector form factor by almost about
20~\%.  On the other hand, the nonlocal terms reduce the
axial-vector form factors almost by 50~\%. It implies that it is
essential to preserve the gauge invariance not only theoretically but
also quantitatively. The large suppression of $F_{A}(Q^2)$ comes
mainly from the nonlocal contributions that are related to diagrams
(b)-(e) in Fig.~\ref{fig:1}.  The contributions from them have been
considered not only by two of the present authors in the same 
model~\cite{Shim:2017wcq} but also by D. G. Dumm et al. in a nonlocal
NJL model~\cite{Dumm:2010hh,GomezDumm:2012qh} for the radiative 
pion decay. We want to emphasize that they are also crucial
to the description of $F_{A}(Q^2)$ for the case of kaon. 
 The nonlocal contribution turns out to be marginal to the
second axial-vector form factor.  

\begin{figure}[htp]
\includegraphics[width = 12cm]{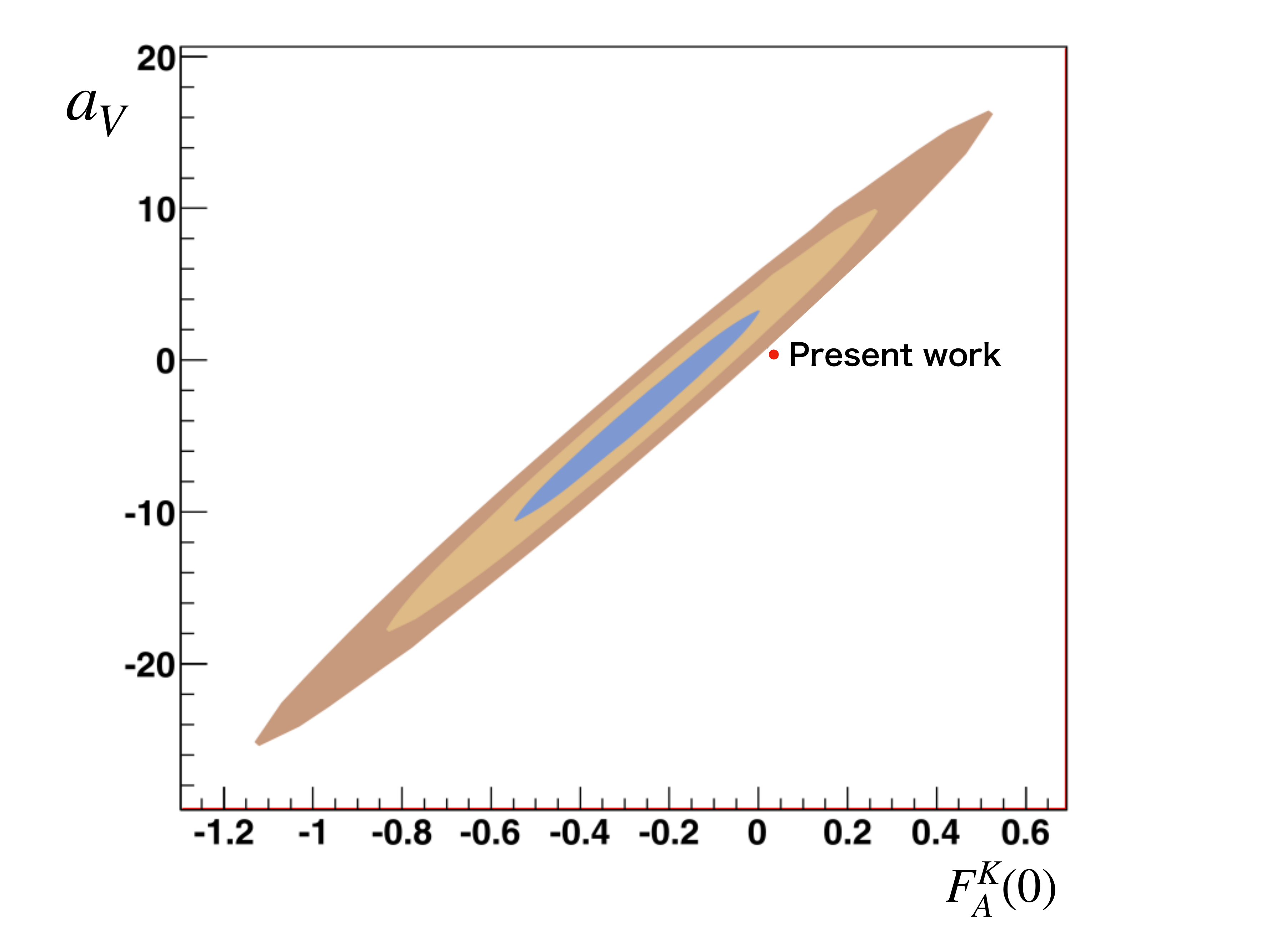} 
\caption{Comparison with the $3\sigma$-ellipse taken from Fig.~22 of
  Ref.~\cite{Duk:2010bs}, where $F_V(0)$ is fixed from
  $\mathcal{O}(p^6)$ $\chi$PT and $a_V$ and $F_A$ are considered as
  fitting parameters. The red blob denotes the present results
  ($a_V=0.383$, $F_A=0.033$).  
  }
\label{fig:4}
\end{figure}
Figure~\ref{fig:4} illustrates the $3\sigma$-ellipse taken from
Fig.~22 of Ref.~\cite{Duk:2010bs}, where $F_V(0)$ is fixed from
$\mathcal{O}(p^6)$ $\chi$PT and $a_V$ and $F_A$ are regarded as
fitting parameters. We can put the present results $a_V=0.404$ and
$F_A=0.027$ as a red blob in Fig.~\ref{fig:4}. Interestingly, the red
blob turns out to be slightly higher than that of $\mathcal{O}(p^6)$
$\chi$PT, which indicates that the present ones are closer to the
$3\sigma$ boundary. However, note that the value of $F_V$ obtained in
the present work is $0.118$ and $0.114$ as shown in Table~\ref{tab:1},
which is almost the same as the experimental data given in
Ref.~\cite{Poblaguev:2002ug}. 

Finally, we extract the parameters for the $p$-pole parametrizations
of the vector and axial-vector form factors for the kaon radiative
decays. In lattice QCD, the $p$-pole parametrization for a 
form factor is often introduced to fit various lattice
data~\cite{Brommel:2006ww, Brommel:2007xd}. Although there is no the
results from lattice QCD yet, it will be useful to 
provide the parameters here so that one can easily compare the present
results with those of lattice QCD in near future. The $p$-pole
parametrizations of the vector and axial-vector form factors are
expressed as 
\begin{align}
F_{V}(Q^2)= \frac{F_{V}(0)}{\left(1+\frac{Q^2}{p_V m_{p_{V}}^2}
  \right)^{p_{V}}},\;\;\; 
F_{A}(Q^2)= \frac{F_{A}(0)}{\left(1+\frac{Q^2}{p_A
  m_{p_{A}}^2}\right)^{p_{A}}},\;\;\; 
R_{A}(Q^2)= \frac{R_{A}(0)}{\left(1+\frac{Q^2}{p_R
  m_{p_{R}}^2}\right)^{p_{R}}}, 
\end{align}
where the results of the parameters $p_V$, $p_A$, $p_R$, $M_{p_A}$,
$M_{p_A}$, and $M_{p_R}$ are listed in Table~\ref{tab:3}. 
\begin{table}[htp]
\centering
\caption{The results of the p-pole parameters. The results listed in
  the row denoted by ``D.P.'' are obtained by using the original
  momentum-dependent quark mass \eqref{eq:ff_dqm} derived in
  Ref.~\cite{Diakonov:1985eg}, whereas those in the last row
  designated by ``Dipole'' are obtained by using the dipole-type form
  factor given in Eq.~\eqref{eq:ff_dipole}.}
\label{tab:3}
\begin{tabular}{m{2cm}m{1cm}m{2cm}m{2cm}m{2cm}m{1cm}m{2cm}}\hline\hline
       & $p_V$  & $M_{p_V}$  & $p_A$ & $M_{p_A}$ & $p_R$ & $M_{p_R}$ \\
       \hline 
Dipole   & $1.26$   & $0.832$ GeV & $1.20$ & $1.15$ GeV &
  $0.779$  & $1.07$ GeV \\
\hline \hline 
\end{tabular}
\end{table}

\section{Summary and conclusion}
In the present work, we investigated the vector and axial-vector
form factors for kaon radiative decays within the framework of the
gauged nonlocal effective chiral action, which constitute the essential
part of the structure-dependent decay amplitude. We scrutinized the
effects of the flavor SU(3) symmetry breaking, releasing the strange
current quark mass from its fixed value $m_{\mathrm{s}}=120$ MeV. The
results showed how the vector and axial-vector form factors undergo
changes when the current quark mass $m$ is varied. We found that the
vector form factor and the second axial-vector form factor increase
monotonically as $m$ increases. On the other hand, the axial-vector
form factor lessens as $m$ increases. 

The numerical results of the form factors are in good agreement with
the experimental data. In general, the experimental data
extracted from the radiative decay of the kaon to the electron are
smaller than those from the radiative pion decay to the muon. The
present results are found to lie between the data taken from the
electron and muon channels. The slope parameter for the axial-vector
form factor was predicted. The $Q^2$ dependences of all the three form
factors were presented and the general tendency is almost the same as
in the case of pion radiative decay. The nonlocal contributions
enhance the vector form factor while they reduce the axial-vector
one. However, their effects are marginal on the second axial-vector
form factors. We compared the present results of the vector slope
parameter and the axial-vector form factor with the $3\sigma$-ellipse
taken from the ISTRA+ Collaboration. Finally, we provided the
parameters for the $p$-pole parametrization of the vector and
axial-vector form factors. 

In the present work, we concentrated only on the vector and
axial-vector form factors for kaon radiative decays. However, it is of
great importance to consider the tensor form factors, though they must
be small experimentally. Since the nonlocal chiral quark model from
the instanton vacuum is a well-defined theoretical framework and
furthermore it does not have any additional free parameter to handle,
it is very interesting to consider the tensor form factors for the
kaon radiative decay. There are at least two important physical
implications on them. Firstly, it offers a possible new physics beyond
the standard model, in particular, related to dark photons. Secondly,
the transition tensor form factors allow one to examine the spin
structure of the kaon in the course of its radiative decay. The
relevant works are under way.  

\section*{Acknowledgments}
H.-Ch. K. is grateful to P. Gubler, T. Maruyama and M. Oka for useful 
discussions.  He wants to express his gratitude to the members of the
Advanced Science Research Center at Japan Atomic Energy Agency for the
hospitality, where part of the present work was done. This work was
supported by the National Research Foundation of Korea (NRF) grant
funded by the Korea government(MSIT) (No. NRF-2018R1A2B2001752).

\end{document}